\documentclass[a4paper]{jpconf}

\usepackage{graphicx}              
\usepackage{amsmath}               
\usepackage{amsfonts}              
\usepackage{amsthm}   		         
\usepackage{color}	
\usepackage{hyperref}	
\usepackage{psfrag}
\usepackage[all, knot]{xy}
\usepackage{calc}

\newcommand\nn{\nonumber}
\newcommand\be{\begin{eqnarray}}
\newcommand\ee{\end{eqnarray}}

\newcommand{\cF}{{\mathcal F}}

\newcommand{\cH}{{\mathcal H}}

\newcommand{\cK}{{\mathcal K}}

\newcommand{\cM}{{\mathcal M}}

\newcommand{\cP}{{\mathcal P}}

\newcommand{\cT}{{\mathcal T}}

\newcommand\U{{\mathrm U}}
\newcommand\C{{\mathbb C}}
\newcommand\N{{\mathbb N}}
\newcommand\R{{\mathbb R}}

\newcommand{\SL}{{\rm SL}}
\newcommand{\SU}{{\rm SU}}

\newcommand{\su}{\mathfrak{su}}

\newcommand{\braaket}[2]{\langle #1  \mid   #2 \rangle }

\newcommand{\braaaket}[3]{\langle #1  \mid   #2   \mid   #3 \rangle }
\newcommand{\bra}[1]{\langle #1  |}
\newcommand{\ket}[1]{| #1 \rangle}
\newcommand{\ketr}[1]{| #1 ]}
\newcommand{\brar}[1]{[ #1 |}

\newcommand{\tz}{\tilde{z}}

\newcommand{\twovec}[2]{\begin{pmatrix} #1 \\ #2 \end{pmatrix}}
\newcommand{\fourmat}[4]{\begin{pmatrix} #1 & #2 \\ #3 & #4 \end{pmatrix}}

\begin{document}

\title{Loop gravity in terms of spinors\footnote{Based on a talk given by one of the authors (JT) at the conference \emph{Loops `11} in Madrid, Spain, on May 27th 2011 \cite{loops11_talk}}}
\author{Etera Livine, Johannes Tambornino}
\address{Laboratoire de Physique, ENS Lyon,\\ CNRS-UMR 5672,\\ 46 All\'ee d'Italie, Lyon 69007, France}
\ead{johannes.tambornino@ens-lyon.fr, etera.livine@ens-lyon.fr}

\begin{abstract}
We show that loop gravity can equally well be formulated in in terms of spinorial variables (instead of the group variables which are commonly used), which have recently been shown to provide a direct link between spin network states and discrete geometries. This results in a new, unitarily equivalent formulation of the theory on a generalized Bargmann space. Since integrals over the group are exchanged for straightforward integrals over the complex plane we expect this formalism to be useful to efficiently organize practical calculations.
\end{abstract}


\section{Introduction and overview}
For technical reasons calculations in loop gravity can in most cases only be performed when truncating the full continuum theory to a single fixed graph. Therefore one needs to know which classical geometrical degrees of freedom are represented by the spin network functions in $\cH_\gamma$, the Hilbert space associated to that graph\footnote{
 In loop gravity the Hilbert space associated to a graph $\gamma$ is given by $\cH_\gamma := L^2(\SU(2)^E, d^Eg)$ where $E$ is the number of edges of that graph and $d^Eg$ the product Haar measure. This space can be interpreted as a quantization of $E$ copies of the cotangent bundle $T^*\SU(2) \simeq \SU(2) \times \su(2)$. $g \in \SU(2)$ is the holonomy of the Ashtekar-connection along an edge $e$ and $X \in \su(2)$ is related to the flux of the densitized triad through a surface dual to $e$. The Hilbert space of the continuum theory arises from the individual graph-Hilbert spaces as an inductive limit
 $
 \cH_{\rm LQG} := \overline{ \cup_\gamma \cH_\gamma / \sim   }
 $,
 where $\sim$ denotes an equivalence relation between states living on different graphs and the completion in an appropriate topology is taken.
}.
As the number of degrees of freedom is always finite, they cannot represent a classical continuum geometry \cite{rovelli_speziale_onegraph}.\\
However, using a parametrization in terms of spinors $\ket{z} \in \C^2$ it was shown in \cite{laurentsimone, laurentsimone2, laurent_etera_un1, laurent_etera_un2} that the classical phase space of loop gravity associated to a graph has a natural interpretation in terms of discrete geometries: in fact, the spin network functions in $\cH_\gamma$ can be seen as a quantum analog of discrete, piecewise flat polyhedral geometries \cite{bianchi_dona_speziale}.\\
This formalism triggered some interesting recent developments: a better understanding of the intertwiner spaces of $\SU(2)$-gauge-invariant loop gravity \cite{etera_spanish_laurent, etera_spanish_un1, etera_spanish_un2} and a new look on the simplicity constraints in spinfoam models \cite{dupuis_livine_simpl1, dupuis_livine_simpl2, dfls_simpl}. See also \cite{livine_tambornino_speziale, wieland_twistors} for a recent generalization of these ideas to the full Lorentz-group. 
\medskip\\
Interestingly it turns out that the use of spinorial variables does not only help to clarify the classical discrete geometry of spin network functions. Also the quantum theory itself can be reformulated exclusively in terms of spinors \cite{livine_tambornino}. The relevant state space $\cH^{\rm spin}_\gamma$ turns out to be a generalization of the Bargmann space \cite{bargmann_reps} of holomorphic square-integrable functions with respect to a Gaussian measure. Thus, in this formulation quantum states of the gravitational field are composed of polynomial functions over the complex numbers instead of functions on $\SU(2)$ as in the standard treatment. We expect this to simplify some computations carried out in loop gravity. One can show that $\cH^{\rm spin}_\gamma$ is unitarily equivalent to $\cH_\gamma$, thus the two formulations contain exactly the same physics. Furthermore, the whole construction is compatible with the inductive limit taken to define the continuum Hilbert space $\cH_{\rm LQG}$, which shows that the spinor techniques do not loose their validity when considering the continuum theory.
\section{The spinorial state space}
The classical phase space associated to the edge-Hilbert space $\cH_e = L^2(\SU(2),dg)$ is the cotangent bundle $T^*\SU(2) \simeq \SU(2) \times \su(2)$, which can be parameterized in terms of a group element $g$, a Lie algebra element $X$ and its `dual' $\tilde{X}= -g^{-1}Xg$.
An alternative parameterization, first used in \cite{laurentsimone}, is the following: consider two spinors $\ket{z}$ and $\ket{\tz}$ living at the initial and final vertex of the edge $e$ respectively\footnote{Our notation is as follows: a spinor $\ket{z} \in \C^2$ has components $\ket{z} := \twovec{z^0}{z^1}$. It has a conjugate $\bra{z} := (\bar{z^0}, \bar{z^1})$ and a dual $\ketr{z} := \epsilon \ket{\bar{z}}, \quad \epsilon := \fourmat{0}{-1}{1}{0}$. The inner product on $\C^2$ is denoted by $\braaket{z}{w} := \bar{z}^0 w^0 + \bar{z}^1w^1$. As a symplectic space $\C^2$ is equipped with the standard symplectic structure $\{ z^A  , \bar{z}^B \} = -i\delta^{AB}, \quad A,B=0,1$.}. From these one can construct two vectors by projecting them on the Pauli matrices as
\be \label{X}
\vec{X}(z) := \braaaket{z}{\vec{\sigma}}{z}, \qquad \vec{\tilde{X}}(\tz) := \braaaket{\tz}{\vec{\sigma}}{\tz} \in \R^3 \, .
\ee
These vectors are then interpreted as oriented areas of the faces of elementary polyhedra living at each vertex.
Furthermore, the following combination turns out to be in the defining representation of $\SU(2)$:
\be \label{g}
g(z,\tz) := \frac{ \ket{z}\brar{\tz} - \ketr{z}\bra{\tz}   }{  \sqrt{\braaket{z}{z} \braaket{\tz}{\tz}}  } 
\ee
Define a constraint that generates $\U(1)$-transformations with opposite sign on both spinors, 
\be \label{constraint}
\cM := \braaket{z}{z} - \braaket{\tz}{\tz} \, .
\ee
Then it can be shown \cite{laurentsimone2} that (excluding some singular points) the symplectic reduction of $\C^2\times\C^2$ with respect to that constraint gives back $T^*\SU(2)$.
\medskip\\
Starting from the spinorial formulation of $T^*\SU(2)$ the most natural Hilbert space to look for a representation of this cotangent bundle is the Bargmann space \cite{bargmann_reps} of holomorphic, square-integrable functions in two complex variables,
\be
\cF_2 := L^2_{\rm hol}(\C^2, d\mu(z)), \quad d\mu(z):= \frac{1}{\pi^2} e^{-\braaket{z}{z}}dz^0 dz^1 \, . \nn
\ee
The space of interest, taking into account the $\U(1)$-constraint (\ref{constraint}), is then
\be
\cH^{\rm spin}_e := \cF_2 \otimes \cF_2 / \U(1) \, . \nn
\ee
The spinors $\ket{z}$ and $\ket{\tz}$ are represented on $\cH^{\rm spin}_e$ as ladder-operators. $g$ and $X$ are then constructed as composite operators via (\ref{X}) and (\ref{g}). 
Restricting attention to $\U(1)$-invariant functions singles out polynomials (labelled by $\alpha, \tilde{\alpha} \in \C \, , \, j \in \frac{1}{2}\N$) of the form 
\be \label{basis}
\cP^j_{\alpha \tilde{\alpha}}(z,\tz) := \frac{1}{(2j)!}\braaket{\alpha}{z}^{2j}\brar{\tz}\epsilon \ket{\tilde{\alpha}}^{2j} \, , 
\ee
which are holomorphic in both spinor variables and further have matching degree. They form an overcomplete basis of $\cH^{\rm spin}_e$, the completeness relations can be derived as
\be 
\int d\mu(z) \int d\mu(\tz) \overline{\cP^j_{\omega \tilde{\omega}}(z,\tz)} \cP^{k}_{\alpha \tilde{\alpha}}(z,\tz) & = & \delta^{jk}\braaket{\alpha}{\omega}^{2j}\braaket{\tilde{\omega}}{\alpha}^{2j} \, ,\nn \\
\sum\limits_j \int d\mu(\omega) d\mu(\tilde{\omega}) \frac{d_j}{(2j)!} \overline{\cP^j_{\omega \tilde{\omega}}(z_1, \tz_1)} \cP^j_{\omega \tilde{\omega}}(z_2, \tz_2) & = & I_0(2\braaket{z_1}{z_2} \braaket{\tz_1}{\tz_2}) \, . \nn
\ee
Here $I_0(x)$ is the zeroth modified Bessel function of first kind and plays the role of the delta-distribution on $\cH^{\rm spin}_e$. These completeness relations are (up to a missing factor of $d_j := 2j+1$ on the right side) exactly the ones fulfilled by the Wigner matrix elements in $L^2(\SU(2),dg)$ when written in the coherent state basis. Thus it is immediate to see that the two spaces are unitarily equivalent. This unitary map can explicitely be written in terms of an integral kernel as
\be \label{map}
\cT_e: && \cH_e \rightarrow \cH^{\rm spin}_e; \\
     && f(g) \mapsto (\cT f)(z, \tz) := \int dg \cK_g(z, \tz) f(g)\, , \nn \\
     && \cK_g(z, \tz) = \sum\limits_{k \in \N} \frac{\sqrt{k+1}}{k!} \brar{\tz}\epsilon g^{-1} \ket{z}^k \, . \nn 
\ee
When applied to Wigner matrix elements in the coherent state basis this map has an interesting interpretation: it essentially (up to some combinatorial factors) restricts the representation matrices of $\SU(2)$, when written in terms of spinors, to their holomorphic part
\be 
 D^j_{\omega \tilde{\omega}}(g) = \left( \bra{\omega} \frac{\ket{z}\brar{\tz} - \ketr{z}\bra{\tz} }{\sqrt{\braaket{z}{z} \braaket{\tz}{\tz}}}  \ket{\tilde{\omega}}  \right)^{2j} \stackrel{\cT}{\mapsto} \frac{1}{(2j)!\sqrt{d_j}}\braaket{\omega}{z}^{2j}\brar{\tz} \epsilon \ket{\tilde{\omega}}^{2j} \, . \nn
\ee
The unitary map (\ref{map}) directly generalizes from a single edge $e$ to an arbitrary graph $\gamma$, showing unitary equivalence between the Hilbert spaces $\cH_\gamma$ and $\cH^{\rm spin}_\gamma = \otimes_e \cH_e^{\rm spin}$:
\be 
\cT_\gamma: \cH_\gamma \rightarrow \cH^{\rm spin}_\gamma \, . \nn
\ee
Thus, equivalence classes of spinor functions living on different graphs $\gamma$ and $\gamma'$ can be defined by demanding the following diagram to commute
\be 
\xymatrix{
  \cH_\gamma \ar[r]^{\cT_{\gamma}} \ar[d]_{\,^*p_{\gamma \gamma'}} & **[r] \cH^{\rm spin}_{\gamma}  \ar[d]^{\,^*p^{\rm spin}_{\gamma \gamma'}}
  \\
  \cH_{\gamma'}    \ar[r]^{\cT_{\gamma'}}   & **[r] \cH^{\rm spin}_{\gamma'}
} \nn
\ee
Here $^*p_{\gamma \gamma'}$ are the isometric embeddings that define equivalence classes on the group side. Their counterparts on the spinor side $^*p_{\gamma \gamma'}^{\rm spin}$ are then used to define equivalence classes of spinor states. Thus, equivalence classes on the left side are mapped to equivalence classes on the right side, no matter which $\cT_\gamma$ is used.  This assures that the construction is cylindrically consistent and allows to abstractly define the \emph{continuum spinor Hilbert space} as
\be
\cH_{\rm LQG}^{\rm spin} := \overline{ \cup_\gamma \cH_\gamma^{\rm spin} / \sim  } \, . \nn 
\ee
Although the exact properties of this space are, for the moment, not very well understood, this shows that the spinor tools can be lifted from a fixed graph to the continuum level.
\medskip\\
One surprising feature of the spinorial formalism, which was discussed in \cite{livine_tambornino}, is that the Haar measure on $\SU(2)$ turns out to be just a Gaussian measure on $\C^4$ when written in terms of spinors, in the sense that 
\be
\int dg f(g) = \int  d\mu(z) \int d\mu(\tz) f(g(z, \tz)) \, , \nn
\ee 
for any $f \in L^2(\SU(2))$ and the group element $g$ interpreted as function of spinors as in (\ref{g}) on the right side. Using spinorial variables to characterize $\SU(2)$ can be understood as choosing a coordinate system with a lot of redundant degrees of freedom. Thus, $f(g(z,\tz))$ is constant along certain directions in $\C^4$ which can be used to turn the Haar measure into Gaussian form. See also \cite{livine_tambornino_speziale} where a similar construction was recently performed for the Haar measure on $\SL(2,\C)$.\\
This Gaussian form of the measure, together with the simple polynomial form of the holomorphic basis (\ref{basis}), is expected to lead to simplification for practical computations: quantities of interest concern the moments of a simple Gaussian measure on $\C^4$ for which combinatorial tools, such as Wick's theorem, are available.
\section{Conclusion and Outlook}
We showed that, based on the recent reformulation of \emph{classical} loop gravity in terms of spinors, one can construct a spinorial Hilbert space that is unitarily equivalent to standard one build over $\SU(2)$. This space is a generalization of the Bargmann space of holomorphic, square-integrable functions over complex numbers. The construction works for an arbitrary graph and is cylincdrically consistent, therefore the lift to the continuum theory is straightforward. Within this new picture quantum states of the gravitational field are represented as holomorphic polynomials over complex numbers, and the measure on the relevant space is of Gaussian form. Detailed calculations in that framework have not been performed yet. But we expect this reformulation, which is much closer to standard field theory than the ordinary version of loop quantum gravity (for example, Wick's theorem directly applies), to push further the development of efficient calculational tools to compute quantities of physical interest (such as scattering amplitudes) within the framework of loop gravity.

\medskip
\ack
This work was partially supported by the ANR ``Programme Blanc'' grant LQG-09. 
\medskip\\

%
\providecommand{\newblock}{}

\end{document}